\documentclass[final,10pt,sort&compress]{elsarticle}
\usepackage{amsfonts,amsmath,amsthm,amssymb,graphicx}
\usepackage{dcolumn,bm,bbding,mathrsfs,graphicx,comment}
\usepackage[colorlinks,citecolor=blue,linkcolor=blue,hyperindex,dvipdfm]{hyperref}

\begin{document}
\begin{frontmatter}
\title{Local quantum uncertainty guarantees the measurement precision for two coupled two-level systems in non-Markovian environment}
\author{Shao-xiong Wu$^1$\corref{cor1}}
\cortext[cor1]{sxwu@nuc.edu.cn}
\author{Yang Zhang$^{2,3}$\corref{}}
\author{Chang-shui Yu$^3$\corref{cor2}}
\cortext[cor2]{quaninformation@sina.com}
\address{$^1$School of Science, North University of China, Taiyuan 030051, China\\
$^2$Department of Physics, Shanxi Datong University, Datong 037009, China\\
$^3$School of Physics, Dalian University of Technology, Dalian 116024, China}
\begin{abstract}
Quantum Fisher information (QFI) is an important feature for the precision of quantum parameter estimation based on the quantum Cram\'{e}r-Rao inequality. When the quantum state satisfies the von Neumann-Landau equation, the local quantum uncertainty (LQU), as a kind of quantum correlation, present in a bipartite mixed state guarantees a lower bound on QFI in the optimal phase estimation protocol [Phys. Rev. Lett. 110 (2013) 240402]. However, in the open quantum systems, there is not an explicit relation between LQU and QFI generally. In this paper, we study the relation between LQU and QFI in open systems which is composed of  two interacting two-level systems coupled to independent non-Markovian environments with the entangled initial state embedded by a phase parameter $\theta$. The analytical calculations show that the QFI does't depend on the phase parameter $\theta$, and its decay can be restrained through enhancing the coupling strength or non-Markovianity. Meanwhile, the LQU is related to the phase parameter $\theta$ and shows plentiful phenomena. In particular, we find that the LQU can well bound the QFI when the coupling between the two systems is switched off or the initial state is Bell state.
\end{abstract}
\end{frontmatter}

\section{Introduction}
Quantum entanglement is a very important physical resource in quantum information \cite{RH09}, recently, it has been shown that the quantum discord \cite{HO01,LH01}, as another kind of quantum correlation, can depict the quantumness of quantum state more deeply than the quantum entanglement in some quantum information processing, such as the DQC1 model \cite{AD08} and is more robust against the decoherence environment. By using different methods, several definitions of quantum correlation were proposed, such as based on quantum information theory \cite{HO01,LH01,Luo08}, Hilbert-Schmidt norm \cite{Dakic10}, relative-entropy \cite{Modi10}, trace norm \cite{FP13} and so on. However, the definition of the quantum correlation that capturing both good physical property and computability is the local quantum uncertainty \cite{DG13}, which is based on the skew information \cite{EW63,SL03prl}.

The local quantum uncertainty (LQU) can not only measure the quantum correlation, but also apply to the field of quantum metrology \cite{DG13,Yu14epl} which was widely used in the field of quantum frequency standard \cite{Santarelli99}, gravitational wave detection \cite{Ligo16}, quantum clock synchronization \cite{VG01} and so on. The goal of parameter estimation is not only to determine the value of the unknown parameters, but also to investigate the accuracy of  the measurements \cite{VG06,VG11}. Just using the classical methods, the limit of the precision of measurements is the standard quantum limit or short noise limit, i.e., $1/\sqrt{N}$, where $N$ is the times of experiments or the numbers of particles used in the experiments \cite{VG11}. However, utilizing the quantum character, such as the quantum squeezing \cite{CC81}, quantum entanglement \cite{SH97}, NOON state \cite{Dorner09}, entangled coherent states \cite{Joo11} and so on, the precision of quantum parameter estimation can be enhanced and surpass the standard quantum limit, even arrive the Heisenberg limit $1/N$. Other quantum technology, such as weak measurement \cite{SP15,LZ15}, dynamical decoupling \cite{QT13}, quantum error correction \cite{WD14}, can also be used to enhance the precision of quantum metrology. The multiple quantum parameters estimation have also been reported \cite{PH13,JY14}.

In the quantum metrology, the precision of quantum parameter estimation is governed by the quantum Cram\'{e}r-Rao inequality $\Delta(\theta)\geq 1/\sqrt{N\mathcal{F}_{\theta}}$, where the quantum Fisher information (QFI) is given by $\mathcal{F}_{\theta}=\mathrm{Tr}\{\rho_{\theta} L_{\theta}^2\}$ with symmetric logarithmic derivative $L_{\theta}$ determined by $2\partial_{\theta}\rho=\rho L_{\theta}+L_{\theta}\rho$. The QFI plays the central role of the quantum metrology, and the larger value of QFI means higher precision of the parameter estimation, when the repeat times $N$ is fixed. In the unitary evolution \cite{VG06}, the probing state can be embedded by a parameter $\theta$ through the unitary transformation, i.e., $\rho_{\theta}=U_{\theta}^{\dagger}\rho U_{\theta}$ with $U_{\theta}=e^{ik\theta}$. The quantum state $\rho_{\theta}$ satisfies the von Neumann-Landau equation $i\partial\rho_{\theta}/\partial\theta=k\rho_{\theta}-\rho_{\theta}k$, the QFI and skew information satisfy the inequality relation \cite{SL03}
\begin{equation}
\mathcal{I}(\rho,k)\leq\frac{1}{4}\mathcal{F}_{\theta}(\rho,k)\leq2\mathcal{I}(\rho,k) .\label{1}
\end{equation}
Through optimizing the operator $k$, the precision of the parameter can be bound by the local quantum uncertainty \cite{DG13}. However, the effect of the environment is not concerned in Refs. \cite{SL03,DG13}. In the open systems, due to the unavoidable interaction with its surroundings \cite{HB07,HB16,AR12}, the unitary dynamics of quantum system will be distorted by the noise, the decoherence effects will influence the precision of the quantum parameter estimation, for example, the metrological advantage using the quantum entanglement will be weaken \cite{Zhangym13} or disappear \cite{SH97}. The quantum metrology in the open systems had been investigated intensively \cite{AC12,XL10,Escher11}. A natural question arises whether the similar relation given in Eq. (\ref{1}) still holds in open systems.

In this paper, we will reexamine the relation between the precision of quantum parameter estimation (QFI) and the quantum correlation (LQU) in the open systems. We employ two coupled two-level systems interacting with the independent non-Markovian environments, respectively; and assume that the initial state with embedded phase parameter $\theta$ is entangled state; and investigate the effects of the environment and the coupling interaction between the two subsystems on the QFI and LQU. We find that the QFI does't depend on the estimated parameter, and its decay can be restrained through enhancing the coupling strength or non-Markovianity. While, the LQU is related to the phase parameter $\theta$ and depicts plentiful phenomena. We find that a similar relation as Eq. (\ref{1}) is satisfied if the coupling between the two systems is switched off or the initial state is Bell state, although there doesn't exist such a bound relation generally. This paper is organized as follows. In Sec. 2, we give the preliminaries about the LQU and QFI. In Sec. 3, the employed model is described. In Sec. 4, the dynamics of the QFI and LQU are investigated, and their relationship is also considered. The conclusion is given in Sec. 5.

\section{The Preliminaries}
In this section, we will give brief introduction on the local quantum uncertainty and the quantum Fisher information. The LQU based on the skew information is given by \cite{DG13}
\begin{eqnarray}
\mathcal{U}(\rho)=\min_{K^{\Lambda}} \mathcal{I}(\rho,K^{\Lambda}), \label{eq:lqu}
\end{eqnarray}
where $K^{\Lambda}=K_{a}\otimes \mathbb{I}_{b}$ means the non-Hermitian
operator on subsystem $A$ with non-degenerate spectrum $\Lambda$. The skew information $\mathcal{I}(\rho,K^{\Lambda})$
denotes the non-commutation between the quantum state $\rho$ and the operator $K^{\Lambda}$, which is defined as
\begin{eqnarray}
\mathcal{I}(\rho,K^{\Lambda})=-\frac{1}{2}\mathrm{Tr}[\rho^{1/2},K^{\Lambda}]^{2}.\label{dingyi:skew}
\end{eqnarray}
For a $2\otimes d$-dimensional system, the LQU can be given by a closed form as
\begin{eqnarray}
\mathcal{U}(\rho)=1-\lambda_{\max}(W_{ab}),\label{lqudingyi}
\end{eqnarray}
where $\lambda_{\max}(\cdot)$ is the maximum eigenvalue and $W_{ab}$ is $3\times3$ symmetric matrix $ (W_{ab})_{ij}=\mathrm{Tr}\{\rho^{1/2}(\sigma_{i}\otimes \mathbb{I}_{b})\rho^{1/2}(\sigma_{j}\otimes \mathbb{I}_{b})\}$ with $i,j=x,y,z$.

The quantum Fisher information is the maximum information about the estimated parameter $\theta$ obtained from optimal measurements. In this paper, the QFI is chosen as the SLD (symmetric logarithmic derivative) definition  $\mathcal{F}_{\theta}(\rho_{\theta})=\mathrm{Tr}\{\rho_{\theta} L_{\theta}^2\}$, where the SLD $L_{\theta}$ is given by $2\partial_{\theta}\rho_{\theta}=\rho_{\theta} L_{\theta}+L_{\theta}\rho_{\theta}$. For quantum state $\rho_{\theta}$, the QFI $\mathcal{F}_{\theta}$ for the estimated parameter $\theta$ is given as follows \cite{MP09,JM11}
\begin{eqnarray}
\mathcal{F}_{\theta}(\rho_{\theta})=\sum_{i}\frac{{(\partial_{\theta}\lambda_{i})^{2}}}{\lambda_{i}}
+\sum_{i\neq j}\frac{{2(\lambda_{i}-\lambda_{j})^{2}}}{\lambda_{i}+\lambda_{j}}
\vert\langle\varphi_{i}\vert\partial_{\theta}\varphi_{j}\rangle\vert^{2},\label{formula:qfi}
\end{eqnarray}
where $\lambda_{i}$ is the eigenvalue of the estimated state $\rho_{\theta}$, $\vert\varphi_{i}\rangle$ is the corresponding eigenvector, and $\partial_{\theta}(\cdot)$ means the partial derivative. For non-full rank density matrix, the expression of the QFI $\mathcal{F}_{\theta}$ can be rewritten as \cite{Zhangym13,LiuJ13,LiuJ14,LiuJ14a}
\begin{eqnarray}
\mathcal{F}_\theta(\rho_\theta)=\sum_{i=1}^r\frac{(\partial_\theta\lambda_i)^2}{\lambda_i}
+\sum_{i=1}^r4\lambda_i\langle\partial_\theta\varphi_i\vert\partial_\theta\varphi_i\rangle
-\sum_{i,j=1}^r\frac{8\lambda_i\lambda_j}{\lambda_i+\lambda_j}\vert\langle\varphi_i\vert\partial_\theta\varphi_j\rangle\vert^2,
\end{eqnarray}
where $r$ is the rank of the density matrix. For pure state $\vert\varphi\rangle$, the QFI can be simplified as $\mathcal{F}_\theta(\vert\varphi\rangle)=4\left(\langle\partial_\theta\varphi\vert\partial_\theta\varphi\rangle
-\vert\langle\varphi\vert\partial_\theta\varphi\rangle\vert^2\right)$.

\section{The model and its solution}
\begin{figure}[t]
  \centering
  \includegraphics[width=0.85\columnwidth]{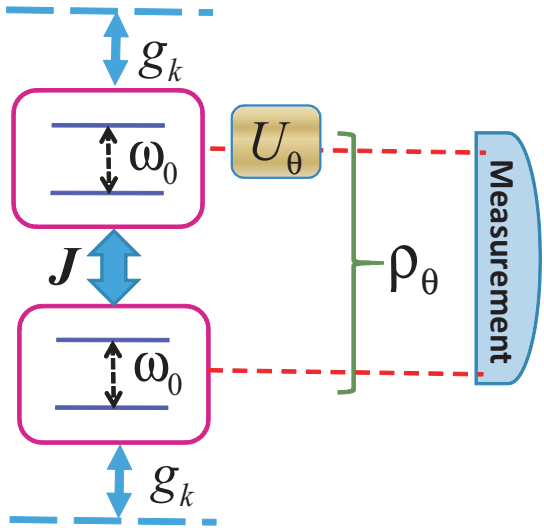}\\
  \caption{(Color online) The model of the phase parameter estimation in open systems. Two coupled atoms $A$ and $B$ interact with non-Markovian environment independently and the coupling strength between the two atoms is $J$. The phase parameter $\theta$ is embedded into the state $\rho_{\theta}$ through unity operation $U_{\theta}$. The quantum correlation and the quantum Fisher information can be obtained after measurement.}\label{tu1}
  \end{figure}

The model for the phase parameter estimation is shown in Fig. \ref{tu1}. The coupled two-level atoms $A$ and $B$ interact with independent non-Markovian environments, respectively. The transition frequency is assumed as the same, i.e., $\omega_{a}=\omega_{b}=\omega_{0}$, the mode of the reservoir is $\omega_{k}$ and the interaction between the subsystem and reservoir is expressed as $g_k$, and the coupling strength between the subsystems $A$ and $B$ is $J$. Due to the interaction between the quantum system and its surroundings, the dynamics of the quantum state $\rho$ consisted of subsystems $A$ and $B$ will be not unitary. The quantum Fisher information and the quantum correlation can be obtained after measurements. In the nature unit, i.e., $\hbar=1$, the whole Hamiltonian for the system and environment is
\begin{equation}
H=H_a+H_b+H_{ab}
\end{equation}
with
\begin{eqnarray*}
H_{a} & = & \omega_{0}\sigma_{a}^{+}\sigma_{a}^{-}+\sum_{k} \omega_{k}a_{k}^{\dagger}a_{k} +\sum_{k}(g_{k}^{*}\sigma_{a}^{+}a_{k}+\mathrm{h.c.}),\\
H_{b} & = & \omega_{0}\sigma_{b}^{+}\sigma_{b}^{-}+\sum_{k}\omega_{k}b_{k}^{\dagger}b_{k} +\sum_{k}(g_{k}^{*}\sigma_{b}^{+}b_{k}+\mathrm{h.c.}),\\
H_{ab} & = & J(\sigma_{a}^{+}\sigma_{b}^{-}+\sigma_{a}^{-}\sigma_{b}^{+}),
\end{eqnarray*}
where, $\sigma_{a}^+(\sigma_b^+)=\vert e\rangle\langle g\vert$ and $\sigma_{a}^-(\sigma_b^-)=\vert g\rangle\langle e\vert$ denote the rasing and lowing operators for subsystem $A(B)$; $a_k^\dag(b_k^\dag)$ and $a_k(b_k)$  represent the creation and annihilation operators for the $k$-th reservoir mode with frequency $\omega_k$ for atom $A(B)$; $H_{ab}$ means the hopping interaction between the subsystems $A$ and $B$. The abbreviation h.c. means the Hermitian conjugation.

Without  loss of generality, we will assume that the initial state of the quantum system is superposition state and both the independent reservoirs interacting with the atoms $A$ and $B$ are vacuum state, i.e., $\vert 0\rangle_a\vert 0\rangle_b$. The initial state of the system and environment is
\begin{equation}
\vert\psi_0\rangle=\left(a_0\vert eg\rangle+b_0\vert ge\rangle\right)\vert0\rangle_{a}\vert0\rangle_{b},
\end{equation}
where $\vert e\rangle$ is the excited state and $\vert g\rangle$ means the ground state, the coefficients $a_0$ and $b_0$ satisfy $\vert a_0\vert^2+\vert b_0\vert^2=1$. Before the evolution, an estimated phase parameter $\theta$ will be embedded into the initial state through unitary operation $U_{\theta}$. The initial state will become $\left(a_0e^{i\theta}\vert eg\rangle+b_0\vert ge\rangle\right)\vert0\rangle_{a}\vert0\rangle_{b}$. In the whole processing, there only exists one excited photon.

In this paper, we only consider that each subsystem resonates with the reservoir mode and the structure of the reservoir is assumed as the Lorentzian form
\begin{eqnarray}
I(\omega)=\frac{1}{2\pi}\frac{\gamma_{0}\lambda^{2}}{\lambda^{2}+(\omega-\omega_{0})^{2}},
\end{eqnarray}
where $\omega_0$ is the transition frequency of the subsystems $A$ and $B$; the parameter $\lambda$ is connected to the reservoir correlation time $\tau_B=1/\lambda$, which is determined by the spectral width of the coupling between the system and reservoir; the parameter $\gamma_0$ is related with the time scale $\tau_R$ of system change $\tau_R=1/\gamma_0$. When $\tau_B\gg\tau_R$ is satisfied, the non-Markovian phenomenon will strongly affect  the dynamics of quantum system.

The evoluted quantum state in time $t$ is
\begin{eqnarray}
\vert\psi(t)\rangle &=& A(t)\vert eg\rangle\vert0\rangle_a\vert0\rangle_b+\sum_{k}C_{k}(t)\vert gg\rangle\vert1_{k}\rangle_a\vert0\rangle_b\notag \\
 &&+B(t)\vert ge\rangle\vert0\rangle_a\vert0\rangle_b+\sum_{k}D_{k}(t)\vert gg\rangle\vert0\rangle_a\vert1_{k}\rangle_b.\label{eq:psi-t}
\end{eqnarray}
Substituting $\vert\psi(t)\rangle$ into the  Schr\"{o}dinger equation $i\vert\dot{\psi}(t)\rangle=H\vert\psi(t)\rangle$, similar with Ref. \cite{HS16}, one can obtain the following differential equations for parameters $A(t),B(t),C_k(t)$ and $D_k(t)$:
\begin{subequations}
\begin{eqnarray}
i\dot{A}(t) & = &\omega_{0}A(t)+J\cdot B(t)+\sum_{k}g_{k}C_{k}(t),\label{eq:At-independent}\\
i\dot{B}(t) & = &\omega_{0}B(t)+J\cdot A(t)+\sum_{k}g_{k}D_{k}(t),\label{eq:Bt-independent}\\
i\dot{C}_{k}(t) & =&  \omega_{k}C_{k}(t)+g_{k}^{*}A(t),\label{eq:ck-independent}\\
i\dot{D}_{k}(t) & =  &\omega_{k}D_{k}(t)+g_{k}^{*}B(t).\label{eq:dk-independent}
\end{eqnarray}\label{A1}
\end{subequations}
The solution of the coefficients $A(t)$ and $B(t)$ can be given as
\begin{eqnarray}
A(t)=a(t)e^{-i\omega_0t},B(t)=b(t)e^{-i\omega_0t}\label{11}
\end{eqnarray}
where
\begin{eqnarray}
a(t)&=&\frac{h}{2}e^{-\frac{1}{2}(\lambda+iJ)t}\left[a_0e^{i\theta}+b_0\right]+\frac{h^{\ast}}{2}e^{-\frac{1}{2}(\lambda-iJ)t}[a_0e^{i\theta}-b_0],\notag\\
b(t)&=&\frac{h}{2}e^{-\frac{1}{2}(\lambda+iJ)t}\left[a_0e^{i\theta}+b_0\right]-\frac{h^{\ast}}{2}e^{-\frac{1}{2}(\lambda-iJ)t}[a_0e^{i\theta}-b_0],\notag\\
h&=&\cosh\left(\frac{d\cdot t}{2}\right)+\frac{\lambda-iJ}{d}\sinh\left(\frac{d\cdot t}{2}\right),\notag\\
d&=&\sqrt{-J^{2}-2iJ\lambda+\lambda(-2\gamma_0+\lambda)}.\label{eq:atbt}
\end{eqnarray}
The detailed derivation is given in the Appendix.

After tracing out the effects of the environment, the reduced density matrix of the quantum system $\rho(t)$ can be given by
\begin{eqnarray}
\rho(t)=\left(\begin{array}{cccc}
0 & 0 & 0 & 0\\
0 & \vert a(t)\vert^{2} & a(t) b(t)^{*} & 0\\
0 & b(t)a(t)^{*} & \vert b(t)\vert^{2} & 0\\
0 & 0 & 0 & 1-\vert a(t)\vert^{2}-\vert b(t)\vert^{2}
\end{array}\right).\label{rho}
\end{eqnarray}
In the following, we will investigate the QFI for the estimated parameter $\theta$, the LQU for the quantum state $\rho(t)$, and the relationship between them.

\section{The dynamics and relationship between QFI and LQU}
\begin{figure}[h!]
  \centering
  \includegraphics[width=1\columnwidth]{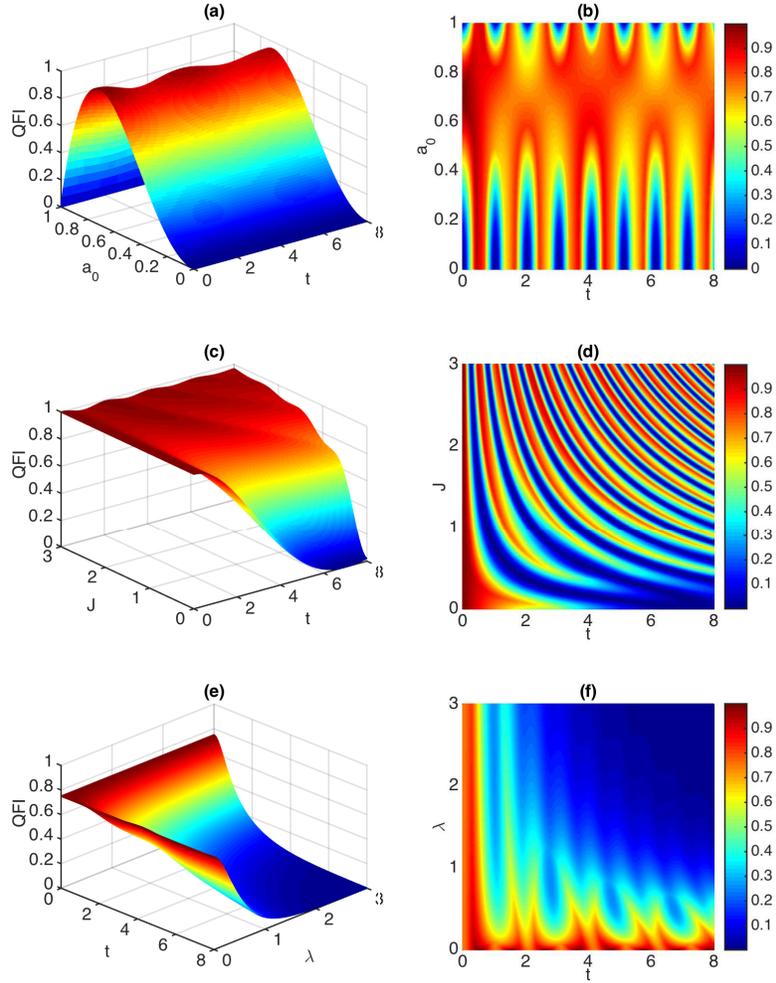}\\
  \caption{The dynamics of the quantum Fisher information and local quantum uncertainty. The variation of QFI and LQU along with the initial state parameter $a_0$ are shown in panels (a) and (b). The parameters are $J=1.5\gamma_0$ (coupling strength), $\lambda=0.1\gamma_0$ (environment parameter) and $\theta=\pi$ (phase parameter). The dynamics of QFI and LQU along with the coupling strength $J$ are shown in panels (c) and (d). The initial state is chosen as Bell state, the other parameters are chosen as $\lambda=0.15\gamma_0$ and $\theta=\pi/2$. The evolution of the QFI and LQU along with the environment parameter $\lambda$ are shown in panels (e) and (f). The other parameters are $J=1.5\gamma_0$, $a_0=0.5$ and $\theta=0$. In all sub-panels, the parameter $\gamma_0$ is chosen as unit.}\label{Fig:1}
\end{figure}

Through tedious calculation, one can obtain that the LQU for quantum state $\rho(t)$ is
\begin{eqnarray}
\mathcal{U}(\rho)=1-\max\{W_1,W_2\},\label{eq:lqu}
\end{eqnarray}
where $W_1,W_2$ are given by
\begin{eqnarray}
W_1=2\vert a(t)\vert^2\sqrt{\frac{\lambda_2}{\lambda_1}},W_2=1-\frac{4\vert a(t)b(t)\vert^2}{\lambda_1},\label{w1w2}
\end{eqnarray}
and $\lambda_i$ is the eigenvalue of quantum state $\rho(t)$ in Eq. (\ref{rho}) with $\lambda_1=\vert a(t)\vert^2+\vert b(t)\vert^2, \lambda_2=1-\lambda_1$.

The QFI for the phase parameter $\theta$ of quantum state $\rho(t)$ can be obtained as follows
\begin{eqnarray}
\mathcal{F}_{\theta}(\rho)=\frac{4\vert b(t)\partial_{\theta}a(t)-a(t)\partial_{\theta}b(t)\vert^2}{\lambda_1}.\label{eq:Fq}
\end{eqnarray}
The amplitude parameters $a(t)$ and $b(t)$ in Eq. (\ref{eq:atbt}) can be rewritten as
\begin{eqnarray}
a(t)&=&\frac{x+x^*}{2}a_0e^{i\theta}+\frac{x-x^*}{2}b_0,\notag\\
b(t)&=&\frac{x-x^*}{2}a_0e^{i\theta}+\frac{x+x^*}{2}b_0,
\end{eqnarray}
where the parameter $x$ is defined as
\begin{eqnarray}
x=e^{-\frac{1}{2}(\lambda+iJ)t} h,\label{eq:x}
\end{eqnarray}
with the parameter $h$ given in Eq. (\ref{eq:atbt}). So, the QFI $\mathcal{F}_{\theta}(\rho)$ in Eq. (\ref{eq:Fq}) can be simplified as
\begin{eqnarray}
\mathcal{F}_{\theta}(\rho)=4\vert a_0b_0\vert^2\vert x\vert^2.\label{eq:Frho}
\end{eqnarray}
One can conclude that the value of QFI does not involve the estimated phase parameter $\theta$.

The dynamics of the QFI and LQU for the reduced system $\rho(t)$ is shown in Fig. \ref{Fig:1}. The evolution of the QFI and LQU along with the initial state parameter $a_0$ and time $t$ are plotted in panels (a) and (b), respectively. The QFI reaches the maximum value in $a_0=1/\sqrt{2}$, i.e., the initial state is Bell state $\vert\psi_0\rangle=\frac{1}{\sqrt{2}}\left(\vert eg\rangle+\vert ge\rangle\right)$, which can also be concluded from Eq. (\ref{eq:Frho}). Unlike the quantum Fisher information, the LQU strongly depends on the phase parameter $\theta$, and exhibits more fertile phenomena than the quantum Fisher information. Because the value of the local quantum uncertainty is determined by the competition between $1-W_1$ and $1-W_2$, the influences suffering from the surrounding environment and the coupling interaction between subsystems will lead to the strongly periodic oscillation in the dynamics of the quantum correlation. In both Fig. \ref{Fig:1} (a) and (b), the coupling strength is $J=1.5\gamma_0$, the non-Makovian parameter is chosen as $\lambda=0.1\gamma_0$. The phase parameter in the dynamics of the quantum correlation is chosen as $\theta=\pi$. Due to the value of the QFI is independent of the estimated phase parameter $\theta$, the phase parameter $\theta$ in panel (a) can be chosen randomly. One can find the similar phenomena when the other parameters condition are chosen.

The evolution of QFI and LQU along with the coupling strength $J$ are shown in Fig. \ref{Fig:1} (c) and (d), where the initial state is chosen as Bell state $\vert\psi_0\rangle=\frac{1}{\sqrt{2}}\left(\vert eg\rangle+\vert ge\rangle\right)$. Because the QFI can reach the maximum once the other condition is fixed. The phase parameter is selected as $\theta=\pi/2$ and the non-Markovian parameter is chosen as $\lambda=0.15\gamma_0$. We can find that the decay of the QFI is rapidly when the coupling interaction $J$ is small. However, as the increasing of the interaction between subsystems, the decay of the QFI can be suppressed dramatically. The coupling interaction $J(\sigma_a^+\sigma_b^-+\sigma_a^-\sigma_b^+)$ can make a hump transition between the quantum state $\vert eg\rangle$ and $\vert ge\rangle$ and the stronger coupling interaction can also make more information about phase $ \theta$ stay at the state $\vert eg\rangle$ or $\vert ge\rangle$, so one can conclude that the strong coupling interaction can significantly restrain the decay of the QFI. However, the LQU displays stronger periodicity as the increasing of coupling strength $J$. It can be explained as the competition between $1-W_1$ and $1-W_2$ can be more intense when the coupling $J$ is bigger. The consequence is that the (quasi-)period of oscillation will be shorter.

In Fig. \ref{Fig:1} (e) and (f), we plot the dynamics of QFI and LQU along with the environment parameter $\lambda$. The initial state coefficient is chosen as $a_0=0.5$, the phase parameter $\theta$ is chosen as $0$, and the coupling strength $J$ is chosen as $1.5\gamma_0$. Due to the interaction between the system and environment, the quantum state $\vert eg\rangle$ or $\vert ge\rangle$ will be transformed into the state $\vert gg\rangle$ and the information of the quantum state will be lost. The parameter $\lambda$ is connected to the spectral width of the reservoir, the small $\lambda$ means not only the weak interaction between the system and reservoir, but also the strong non-Markovianity. So the process of losing information will slow down, and the strong non-Markovianity means the information can backflow from the environment to the quantum system \cite{XL10}. So, one can find that the decay of the QFI and LQU can be suppressed  when the environment parameter $\lambda$ is smaller, in other word, the non-Markovianity is more remarkable.

When the quantum state $\rho_{\theta}$ satisfies the von Neumann-Landau equation $i\partial\rho_{\theta}/\partial\theta=k\rho_{\theta}-\rho_{\theta}k$ ($\rho_\theta$ can be generated by operator $k$ through $\rho_{\theta}=e^{-i\theta k}\rho e^{i\theta k}$), the quantum Fisher information and the skew information satisfy the inequality
\begin{eqnarray}
\mathcal{I}(\rho,k)\leq\frac{1}{4}\mathcal{F}_{\theta}(\rho,k)\leq2\mathcal{I}(\rho,k).\label{ineq:luo}
\end{eqnarray}
For two-qubits system, if the operator $k$ is the local observable on subsystem $A$ , such as $k=K_{a}\otimes \mathbb{I}_{b}$, optimizing all local observable $k$, one can arrive the definition of local quantum uncertainty in Eq. (\ref{lqudingyi}). Using the quantum Cram\'{e}r-Rao inequality $\Delta(\theta)\geq 1/\sqrt{N\mathcal{F}_{\theta}}$, the parameter precision can be bound by LQU $\Delta(\theta)<1/\sqrt{4\mathcal{U}}$ when the repeat number is $N=1$ \cite{DG13}. A nature question is whether the QFI and LQU still satisfy the similar relationship in the open quantum systems, for example the model depicted in Fig. \ref{tu1}.

\begin{figure}[t]
  \centering
  \includegraphics[width=1\columnwidth]{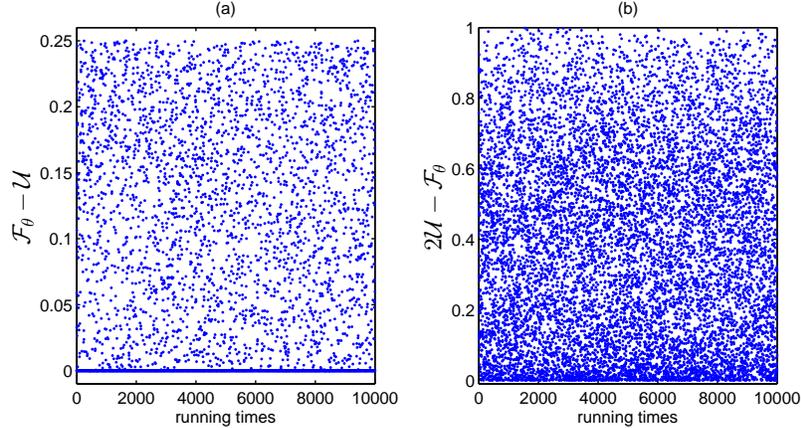}\\
  \caption{(Color online) The difference between the QFI and LQU. The coupling strength $J$ is $0$, the parameter $\gamma_0$ is chosen as unit, and the other parameters are chosen randomly. The number of running times is $10^4$.} \label{Fig3}
\end{figure}

With the help of numerical simulation, one can see that the QFI and LQU do not satisfy the inequality relation (\ref{ineq:luo}) generally. However, for some special cases, one can also find that the QFI is bounded by the LQU. In Fig. \ref{Fig3}, the difference between QFI and LQU is dotted $10^4$ times, where the coupling strength is $J=0$ and the other parameters are chosen randomly. The difference of $\mathcal{F}_{\theta}-\mathcal{U}$ is shown in panel (a), and the maximum difference is $1/4$ and the minimum value is $0$. The difference of $2\mathcal{U}-\mathcal{F}_{\theta}$ is shown in panel (b) and its range is from $0$ to $1$. This can be given a simple proof.

When the coupling strength $J$ between the two subsystems is $0$, in other worlds, the coupling between the two systems is switched off, the parameter $x$ in Eq. (\ref{eq:x}) can be simplified as
\begin{eqnarray}
x_0=e^{-\frac{1}{2}\lambda t} h_0,
\end{eqnarray}
where $h_0=\cosh(d_0t/2)+\lambda/d_0 \sinh(d_0t/2)$ with $d_0= \sqrt{\lambda(-2\gamma_0+\lambda)}$. It is easy to find that the parameter $x_0$ is real, and the amplitude parameters $a(t)$ and $b(t)$ can be simplified as
\begin{eqnarray}
a(t)=a_0e^{i\theta}x_0,b(t)=b_0x_0.
\end{eqnarray}
The eigenvalues $\lambda_1$ and $\lambda_2$ for the reduced density matrix of quantum state $\rho(t)$ in Eq. (\ref{rho}) can be simplified as $\lambda_1=\vert x_0\vert^2$ and $\lambda_2=1-\vert x_0\vert^2$.

Firstly, we will prove that the value of LQU is smaller than the value of QFI. The QFI $\mathcal{F}_{\theta}(\rho)$ in Eq. (\ref{eq:Frho}) can be simplified as
\begin{eqnarray}
\mathcal{F}_{\theta}(\rho)=4\vert a_0b_0\vert^2\vert x_0\vert^2.
\end{eqnarray}
The LQU $\mathcal{U}(\rho)$ in Eq. (\ref{eq:lqu}) can be rewritten as
 \begin{eqnarray}
 \mathcal{U}(\rho)=\min\{\mathcal{U}_1,\mathcal{U}_2\},
 \end{eqnarray}
 where $\mathcal{U}_1, \mathcal{U}_2$ are given by
 \begin{eqnarray}
 \mathcal{U}_1=1-2\vert a_0\vert^2\sqrt{\vert x_0\vert^2(1-\vert x_0\vert^2)}, \mathcal{U}_2=4\vert a_0b_0\vert^2\vert x_0\vert^2.
 \end{eqnarray}
Obviously, the value of $\mathcal{F}_{\theta}$ equals with $\mathcal{U}_2$. The value of LQU is always chosen the smaller one between $\mathcal{U}_1$ and $\mathcal{U}_2$, so the value of LQU is $\mathcal{U}_1$ if and only if $\mathcal{U}_1<\mathcal{U}_2$, otherwise the value of LQU will be determined by $\mathcal{U}_2$. So, the LQU $\mathcal{U}$ and QFI $\mathcal{F}_{\theta}$ meet the following relation:
\begin{eqnarray}
\mathcal{U}\leq\mathcal{F}_{\theta}.\label{ineq1}
\end{eqnarray}

In the following, we will find out the maximum difference between the QFI and LQU. Due to the parameter $\mathcal{U}_2$ equals to the value of $\mathcal{F}_{\theta}$, so we only pay attention to the condition that the value of LQU is chosen as $\mathcal{U}_1$. For simplicity, we will assume that $\vert a_0\vert^2=m$, $\vert x_0\vert^2=n$, and the range of $m$ and $n$ are both from $0$ to $1$. The difference between the QFI and LQU can be expressed as
\begin{eqnarray}
\delta_1&=&{\mathcal{F}_{\theta}-\mathcal{U}}\notag\\
&=&4m(1-m)n-\left(1-2m\sqrt{n(1-n)}\right).\label{qfiminuslqu}
\end{eqnarray}
The second derivative of $m$ is negative, i.e., $\partial^2\delta_1/\partial m^2<0$, so the function of $\delta_1$ reaches the maximum value at the point $\partial\delta_1/\partial m=0$. After some algebra, one can obtain that
\begin{eqnarray}
\max\delta_1=\frac{1}{4},
\end{eqnarray}
which is displayed in the Fig. \ref{Fig3}(a).

Then, we will prove that $\mathcal{F}_{\theta}\leq2\mathcal{U}$. The value of LQU is also chosen as $\mathcal{U}_1$, the function of $2\mathcal{U}-\mathcal{F}_{\theta}$ can be expressed as
\begin{eqnarray}
\delta_2&=&2\mathcal{U}-\mathcal{F}_{\theta}\notag\\
&=&2\left(1-2m\sqrt{n(1-n)}\right)-4m(1-m)n.
\end{eqnarray}
One can obtain that the second derivative of $\delta_2$ is positive ($\partial^2\delta_2/\partial m^2>0$), so the function of $\delta_2$ arrives the minimum value at the point $\partial\delta_2/\partial m=0$ and $\min\delta_2=\left(\sqrt{1-n}-\sqrt{n}\right)^2\geqslant0$. It will lead the following inequality
\begin{eqnarray}
\mathcal{F}_{\theta}\leq2\mathcal{U}.\label{ineq2}
\end{eqnarray}
Combining Eqs. (\ref{ineq1}) and (\ref{ineq2}), one can conclude that the QFI and LQU still satisfy the following relation:
\begin{eqnarray}
\mathcal{U}(\rho)\leq\mathcal{F}_{\theta}(\rho)\leq2\mathcal{U}(\rho),\label{eq23}
\end{eqnarray}
when the two subsystems have no interaction.

In another case, the initial state is chosen as Bell state $\vert\psi_0\rangle=\frac{1}{\sqrt{2}}(\vert eg\rangle+\vert ge\rangle)$, i.e., $a_0=b_0=1/\sqrt{2}$. The QFI of phase parameter $\theta$ is
\begin{eqnarray}
\mathcal{F}_{\theta}(\rho)=\vert x\vert^2,
\end{eqnarray}
where the parameter $x$ is given in Eq. (\ref{eq:x}). The LQU for the quantum state can be expressed as
\begin{eqnarray}
\mathcal{U}(\rho)=\min\{\mathcal{U}_1,\mathcal{U}_2\},
\end{eqnarray}
where
\begin{eqnarray}
\mathcal{U}_1=1-\frac{2\vert a(t)\vert^2\sqrt{\vert x\vert^2(1-\vert x\vert^2)}}{\vert x \vert^2},
\mathcal{U}_2=\frac{4\vert a(t)\vert^2(\vert x\vert^2-\vert a(t)\vert^2)}{\vert x\vert^2},\notag
\end{eqnarray}
with $\vert a(t)\vert^2=\vert x\vert^2/2+\mathfrak{R}\mathfrak{I}\sin\theta$, $\mathfrak{R}$ meaning the real part of $x$ and $\mathfrak{I}$ meaning the image part. Obviously, the value of $\mathcal{U}_2$ and $\mathcal{F}_{\theta}$  satisfy the inequality $\mathcal{U}_2\leq\mathcal{F}_{\theta}$. So, when the value of LQU is chosen as $\mathcal{U}_2$, one can conclude that $\mathcal{U}\leq\mathcal{F}_{\theta}$. In the condition that the value of $\mathcal{U}_1$ is smaller than $\mathcal{U}_2$, the value of LQU will be chosen as $\mathcal{U}_1$, the inequality $\mathcal{U}\leq\mathcal{F}_{\theta}$ is satisfied automatically. However, the QFI and LQU does't satisfy the relationship $\mathcal{F}_{\theta}(\rho)\leq 2\mathcal{U}(\rho)$ generally, which can be obtained with the help of numerical simulation.

When the state satisfies the von Neumann-Landau equation, the amount of the quantum correlation present in a bipartite mixed state guarantees a minimum precision in the optimal phase estimation protocol \cite{DG13}. However, considering the effects of the environment and the coupling interaction between subsystems, the quantum Fisher information and the local quantum uncertainty don't satisfy the similar relationship generally. However, when the coupling strength between the two subsystems is zero or the initial state is Bell state, we can also obtain the inequality $\mathcal{U}(\rho)\leq\mathcal{F}_{\theta}(\rho)$, the optimal detection strategy which asymptotically saturates the quantum Cram\'{e}r-Rao bound produces the best precision of parameter $\theta _{\mathrm{best}}$
\begin{eqnarray}
\Delta({\theta _{\mathrm{best}}}) = \frac{1}{\sqrt{\mathcal{F}_{\theta}}}\le\frac{1}{\sqrt{ \mathcal{U}}}.
\end{eqnarray}
Obviously, the quantum parameter precision can still be bound by the quantum correlation in the above two special cases.

\section{Conclusion}

In summary, the local quantum uncertainty is not only a good measurement of quantum correlation, but also an effective bound on the precision of parameter estimation through the relation between the quantum Fisher information and skew information. In this paper, we reexamined the relationship between the local quantum uncertainty and the precision of the parameter estimation through extending the parameter estimation protocol to the open quantum systems. We employ the coupled two-level systems interacting with independent non-Markovian reservoir, in which the initial state is entangled state with embedded phase parameter $\theta$ through unitary operation. In general, the precision of phase parameter can not be bound by the quantum correlation. However, for some special cases, for example the coupling between two subsystems is switched off or the initial state is Bell state, the phase parameter precision can also be bound by the quantum correlation. In this paper, we only investigate the structure of environment is Lorentzian form, the general relationship between the QFI and LQU in the open systems is deserved endeavor in our further investigation.

\section*{Acknowledgments}
This work was supported by the National Natural Science Foundation of China under Grants No.11747022, 11775040 and 11375036, the Xinghai Scholar Cultivation Plan, the Doctoral Startup Foundation of North University of China (No.130088), and the Science Foundation of North University of China (No.2017031).

\renewcommand{\theequation}{A\arabic{equation}}
\setcounter{equation}{0}
\section*{Appendix:  The derivation of $a(t)$ and $b(t)$ in Eq. (\ref{eq:atbt})}

In order to solve the equations (\ref{A1}), the frame rotating is used: $A(t)=a(t)e^{-i\omega_{0}t}$, $B(t)=b(t)e^{-i\omega_{0}t}$, $C_{k}(t)=c_{k}(t)e^{-i\omega_{k}t}$, $D_{k}(t)=d_{k}(t)e^{-i\omega_{k}t}$. For the Eq. (\ref{eq:ck-independent}), one can obtain that
\begin{eqnarray}
\dot{c}_{k}(t)=-i g_{k}^{*}a(t)e^{-i(\omega_{0}-\omega_{k})t}.
\end{eqnarray}
The form integral for $c_k(t)$ is
\begin{equation}
c_{k}(t)=-i\int_{0}^{t}g_{k}^{*}a(\tau)e^{-i(\omega_{0}-\omega_{k})\tau}\text{d}\tau. \label{eq:ck-jifen}
\end{equation}

Replacing the Eq. (\ref{eq:ck-jifen}) into Eq. (\ref{eq:At-independent}), one can obtain the following integro-differential equation:
\begin{eqnarray}
\dot{a}(t)=-iJ\cdot b(t)-\sum_{k}|g_{k}|^{2}\int_{0}^{t}a(\tau)e^{-i(\omega_{0}-\omega_{k})(t-\tau)}\text{d}\tau.
\end{eqnarray}
Assuming the coupling structure between the atom and reservoir is $I(\omega)$ and taking the limit of reservoir mode, one can obtain the final integro-differential equation for $a(t)$
\begin{eqnarray}
\dot{a}(t)=-iJ\cdot b(t)-\int_{0}^{t}\text{d}\tau f(t-\tau)a(\tau),\label{A5}
\end{eqnarray}
where the kernel $f(t-\tau)$ can be expressed in terms of the spectral density of reservoir $I(\omega)$
\begin{equation}
f(t-\tau)=\int\text{d}\omega I(\omega)e^{i(\omega_{0}-\omega)(t-\tau)}.    \label{eq:fttau}
\end{equation}
Similarly, one can arrive the integro-differential equation for $b(t)$
\begin{eqnarray}
\dot{b}(t)=-iJ\cdot a(t)-\int_{0}^{t}\text{d}\tau f(t-\tau)b(\tau).\label{A7}
\end{eqnarray}

In order to solve the integro-differential equation (\ref{A5}) and (\ref{A7}), the spectral density of reservoir is assumed as the Lorentzian form
\[
I(\omega)=\frac{{1}}{2\pi}\frac{\gamma_{0}\lambda^{2}}{\lambda^{2}+(\omega-\omega_{0})^{2}}.
\]
Using the Laplace transformation $F(s)=\int_0^{\infty}f(t)e^{-st}\text{d}t$, the integro-differential equations (\ref{A5}) and (\ref{A7}) can be solved as
\begin{eqnarray}
a(s)&=&\frac{-iJ\cdot b_0+[s+f(s)]a_0e^{i\theta}}{J^{2}+[s+f(s)]^{2}},\notag\\
b(s)& = & \frac{-iJ\cdot a_0e^{i\theta}+[s+f(s)]b_0}{J^{2}+[s+f(s)]^{2}}.\label{Abs}
\end{eqnarray}
In order to obtain the amplitude parameters $a(t)$ and $b(t)$ in Eq. (\ref{eq:atbt}), the inverse Laplace transform for Eq. (\ref{Abs}) is used, i.e., $f(t)=\frac{{1}}{2\pi i}\int_{\gamma-i\infty}^{\gamma+i\infty}F(s)e^{st}\text{d}s$
(positive arbitrary constant $\gamma$ are chosen lied to the
right of all the singularities of functions $F(s)$).


\begin{thebibliography}{99}
\bibitem{RH09} R. Horodecki, P. Horodecki, M. Horodecki and K. Horodecki. Rev. Mod. Phys. 81(2009) 865.
\bibitem{HO01} H. Ollivier and W. H. Zurek. Phys. Rev. Lett. 88 (2001) 017901.
\bibitem{LH01} L. Henderson and V. Vedral. J. Phys. A 34 (2001) 6899.
\bibitem{AD08} A. Datta, A. Shaji, and C.M. Caves, Phys. Rev. Lett. 100 (2008) 050502.

\bibitem{Luo08} S. Luo. Phys. Rev. A 77 (2008) 042303.
\bibitem {Dakic10} B. Daki\'{c}, V. Vedral and C. Brukner. Phys. Rev. Lett. 105 (2010) 190502.
\bibitem{Modi10} K. Modi, T. Paterek, W. Son, V Vedral, and M. Williamson. Phys. Rev. Lett. 104 (2010) 080501.
\bibitem{FP13} F. M. Paula,T. R. de Oliveira, and M. S. Sarandy. Phys. Rev. A 87 (2013) 064101.
\bibitem{DG13} D. Girolami, T. Tufarelli, and G. Adesso, Phys. Rev. Lett. 110 (2013) 240402.
\bibitem{EW63} E. P. Wigner and M. M. Yanase. Proc. Natl. Acad. Sci.  49 (1963) 910.
\bibitem{SL03prl} S. Luo. Phys. Rev. Lett. 91 (2003) 180403.
\bibitem{Yu14epl} C. S. Yu, S. X. Wu, X. Wang, X. X. Yi and H. S. Song. EPL. 107 (2014) 10007.
\bibitem{Santarelli99} G. Santarelli, Ph. Laurent, P. Lemonde, et al. Phys. Rev. Lett. 82 (1999) 4619.
\bibitem{Ligo16} The LIGO Scientific Collaboration. Phys. Rev. Lett. 116 (2016) 061102.
\bibitem{VG01} V. Giovannetti, S. Lloyd and L. Maccone. Nature 412 (2001) 417.
\bibitem{VG06} V. Giovannetti, S. Lloyd, and L. Maccone, Phys. Rev. Lett. 96 (2006) 010401.
\bibitem{VG11} V. Giovannetti, S. Lloyd and L. Maccone. Nat. Photonics 5 (2011) 222.
\bibitem{CC81} C. M. Caves. Phys. Rev. D 23 (1981) 1693.
\bibitem{SH97} S. F. Huelga, C. Macchiavello, T. Pellizzari, A. K. Ekert, M. B. Plenio, J. I. Cirac. Phys. Rev. Lett. 79 (1997) 3865.
\bibitem{Dorner09} U. Dorner, R. Demkowicz-Dobrzanski, B. J. Smith, et al. Phys. Rev. Lett. 102 (2009) 040403.
\bibitem{Joo11} J. Joo, W. J. Munro, and T. P. Spiller. Phys. Rev. Lett. 107 (2011) 083601.
\bibitem{SP15} S. Pang and T. A. Brun, Phys. Rev. Lett. 115 (2015) 120401.
\bibitem{LZ15} L. Zhang, A. Datta, and I. A. Walmsley, Phys. Rev. Lett. 114 (2015) 210801.
\bibitem{QT13} Q. S. Tan, Y. Huang, X. Yin, L. Kuang, and X. Wang. Phys. Rev. A 87 (2013) 032102.
\bibitem{WD14} W. D\"{u}r, M. Skotiniotis, F. Fr\"{o}wis, and B. Kraus. Phys. Rev. Lett. 112 (2014) 080801.
\bibitem{PH13} P. C. Humphreys, M. Barbieri, A. Datta, and Ian A. Walmsley. Phys. Rev. Lett. 111 (2013) 070403.
\bibitem{JY14} J. D. Yue, Y. R. Zhang and H. Fan. Sci. Rep. 4 (2014) 5933.
\bibitem{SL03} S. Luo. Pro. Amer. Math. Soc. 132 (2003) 885.
\bibitem{HB07} H. P. Breuer and F. Petruccione. The theory of open quantum systems. New York: Oxford University Press, (2007).
\bibitem{HB16} H. P. Breuer, E. M. Laine, J. Piilo, and B. Vacchini. Rev. Mod. Phys. 88 (2016) 021002.
\bibitem{AR12} \'{A}. Rivas and S. F. Huelga. Open quantum systems (Springer, 2012).
\bibitem{Zhangym13} Y. M. Zhang, X. W. Li, W. Yang, and G. R. Jin. Phys. Rev. A 88 (2013) 043832.
\bibitem{Escher11} B. M. Escher, R. L. de Matos Filho and L. Davidovich.  Nat. Phys. 7 (2011) 406.
\bibitem{XL10} X. M. Lu, X. Wang, and C. P. Sun. Phys. Rev. A 82 (2010) 042103.
\bibitem{AC12} A. W. Chin, S. F. Huelga, and M. B. Plenio. Phys. Rev. Lett. 109 (2012) 233601.
\bibitem{MP09} M. G. A. Paris. 	Inter. J. Quantum Infor. 7 (2009) 125.
\bibitem{JM11} J. Ma, X. Wang, C. P. Sun and F. Nori. Phys. Rep. 509 (2011) 89.
\bibitem{LiuJ14} J. Liu, X. X. Jing, W. Zhong, and X. Wang. Commun. Theor. Phys. 61 (2014) 45.
\bibitem{LiuJ13} J. Liu, X. Jing, and X. Wang. Phys. Rev. A 88 (2013) 042316.
\bibitem{LiuJ14a} J. Liu, H. N. Xiong, F. Song, X. Wang. Physica A 410 (2014) 167.
\bibitem{HS16} H. Z. Shen, X. Q. Shao, G. C. Wang, X. L. Zhao, and X. X. Yi. Phys. Rev. E 93 (2016) 012107.

\end{thebibliography}
\end{document}